\documentclass{iaus}
\usepackage{graphicx}

\title[JD 11.~~Modelling colliding-wind binaries] 
{Non-thermal radio emission from colliding-wind binaries: \\ modelling Cyg~OB2 No.~8A and  No.~9
}

\author[Delia Volpi, Ronny Blomme, Michael De Becker, \& Gregor Rauw]   
{Delia Volpi$^1$, Ronny Blomme$^1$, Michael De Becker$^2$$^{,3}$, \and Gregor Rauw$^2$
}

\affiliation{$^1$Royal Observatory of Belgium, Ringlaan 3, B-1180 Brussels, Belgium, \\ email: {\tt delia.volpi@oma.be}\\[\affilskip]
$^2$Institut d'Astrophysique, Universit\'e de Li$\mathrm{\grave{e}}$ge,\\ All\'ee du 6 Ao$\mathrm{\hat{u}}$t, 17, B$\mathrm{\hat{a}}$t B5c, B-4000 Li$\mathrm{\grave{e}}$ge (Sart-Tilman), Belgium\\[\affilskip]
$^3$Observatoire de Haute-Provence, F-04870 Saint-Michel l'Observatoire, France}

\pubyear{2010}
\volume{272}  
\pagerange{1-2}
\jname{Active OB stars: structure, evolution, mass loss and critical limits}
\editors{C.Neiner, G. Wade, G. Meynet \& G. Peters, eds.}

\begin{document}

\maketitle

\begin{abstract} 
Some OB stars show variable non-thermal radio emission. The non-thermal emission is due to synchrotron radiation that is emitted by electrons accelerated to high energies. The electron acceleration occurs at strong shocks created by the collision of radiatively-driven stellar winds in binary systems. Here we present results of our modelling of two colliding wind systems: Cyg~OB2 No.~8A and Cyg~OB2 No.~9.

\keywords{plasmas, radiation mechanisms: nonthermal, methods: numerical, binaries: spectroscopic, stars: early-type, stars: winds, outflows}
\end{abstract}

\firstsection 
\section{Introduction}

\begin{figure}[t]
\begin{center}
\includegraphics[width=2.0in]{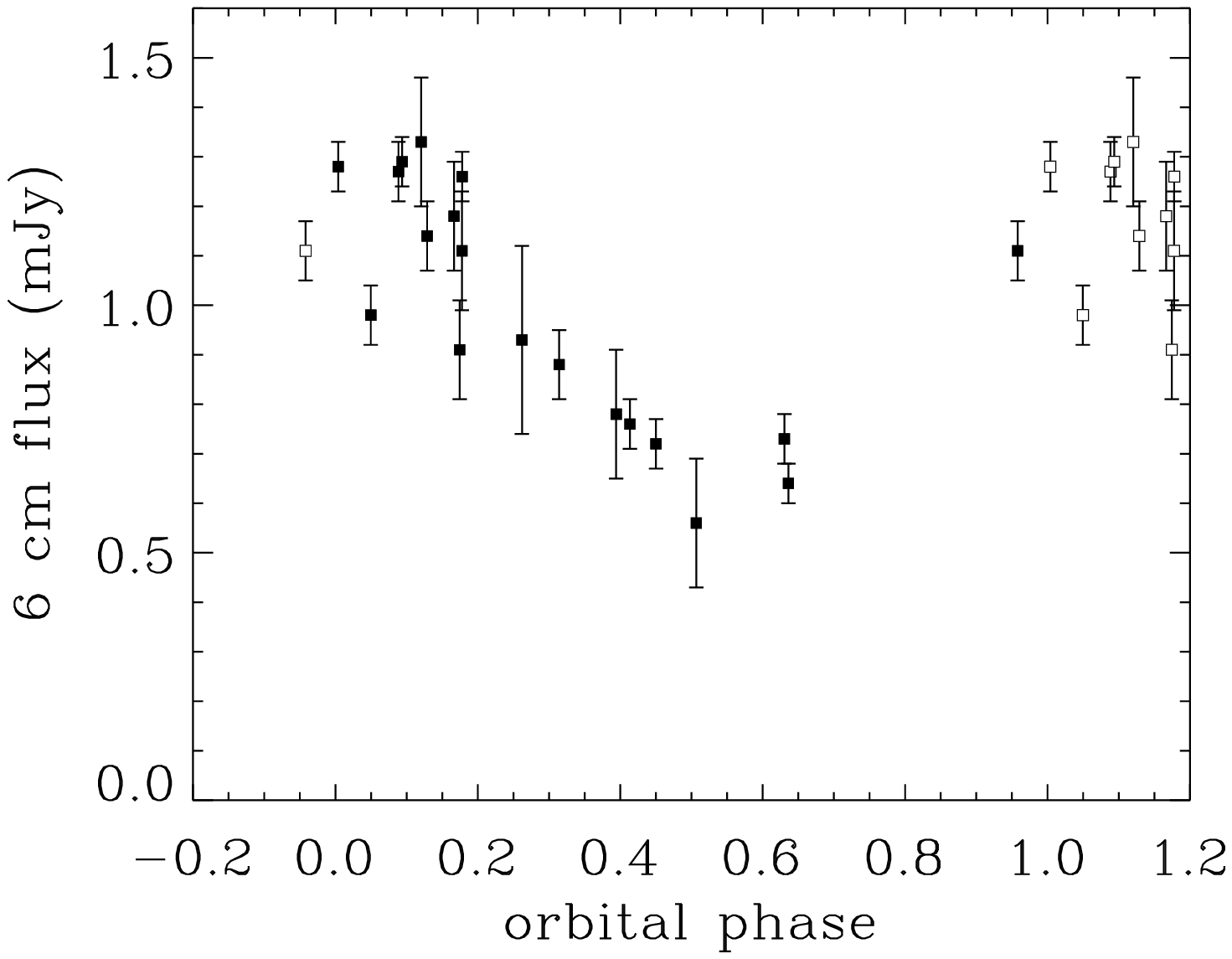} 
\includegraphics[width=2.0in]{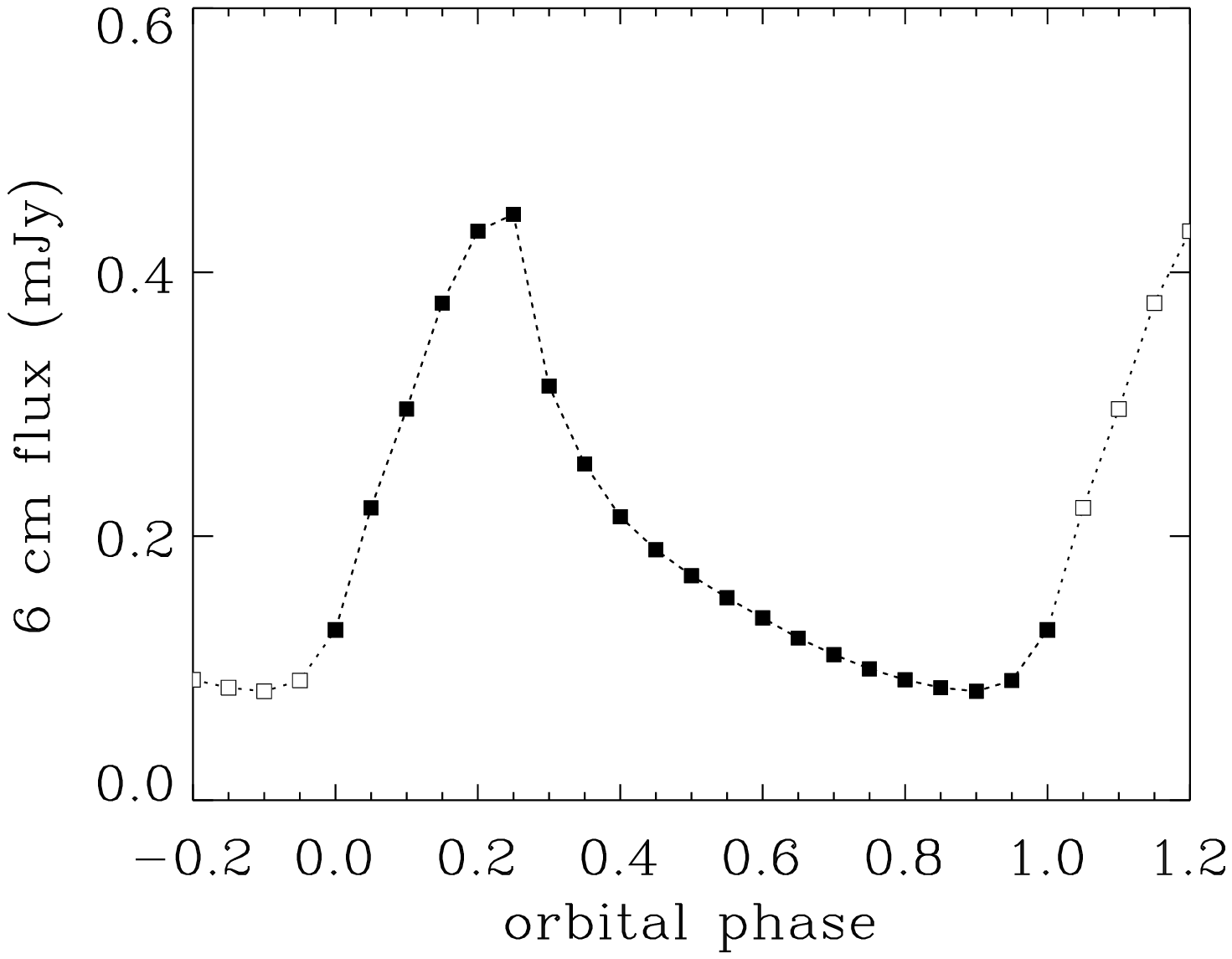} 
\caption{Flux in the radio band at 6 cm for Cyg~OB2 No.~8A: on the left the observations from \cite[Blomme et al. (2010)]{Blomme_etal10}, on the right our simulated results. Periastron is at phase $\approx 0$}
\label{fig1}
\end{center}
\end{figure}
During recent years many OB stars have been discovered to be binary systems. Non-thermal radio emission is observed to be produced by some of these binary stars. The non-thermal emissivity is thought to be due to synchrotron emission radiated by relativistic electrons. The electrons are accelerated up to high energies by strong shocks produced by the collision between the two radiatively driven stellar winds (\cite[Eichler \& Usov 1993]{EichlerUsov93}). Several parameters of the system can be constrained by the synchrotron emission, among them the mass loss rates from the primary and the secondary. Investigating the synchrotron radiation is thus necessary. We model the non-thermal emission for two colliding wind systems, Cyg~OB2 No.~8A and Cyg~OB2 No.~9, and compare the obtained results with the observations. 

\section{Modelling and comparison with the data}
The two colliding winds are separated by a contact discontinuity. Its position (which, in our model, we assume to be coincident with the two shocks) is defined as in \cite[Antokhin et al. (2004)]{Antokhin_etal04}.
The electrons are accelerated at the shock. We follow them as they advect away and cool down due to adiabatic and inverse Compton losses along the post-shock streamlines. The momenta follow a modified power-law distribution. The synchrotron emissivity from the relativistic electrons is calculated along the post-shock streamlines in the orbital plane. The Razin effect is included. The third dimension is recovered by rotating the orbital plane along the line which connects the two stars. We also include free-free emission and then calculate the fluxes and spectral indices at different orbital phases using Adam’s method (\cite{Adam90}). For Cyg ~OB2 No.~8A the parameters are provided by \cite[De Becker et al. (2006)]{DeBecker_etal06}, for Cyg~OB2 No.~9 the orbital parameters are provided by \cite[Naz\'e et al. (2010)]{Naze_etal10}, stellar parameters by \cite[Martins et al. (2005)]{Martins_etal05} and wind parameters by \cite[Vink et al. (2001)]{Vink_etal01}.

\paragraph{\bf Cyg~OB2 No.~8A results.}
For Cyg~OB2 No.~8A (see \cite{Blomme_etal10}) the radio data at 3.6 and 6 cm are obtained with VLA, the X-ray data with XMM and ROSAT. Variability that is locked with the orbital phase is observed in both radio and X-rays. The X-ray and the radio light curves are anti-correlated due to different formation regions. The radio formation region is far out in the wind, along the contact discontinuity, while the X-rays are formed much closer to the apex of the contact discontinuity. The model predicts phase-locked radio variability which is consistent with the observations (see Fig.~\ref{fig1}), even if the phases of the flux maximum and minimum do not agree. 

\paragraph{\bf Cyg~OB2 No.~9 results.}
\cite{vanLoo_etal08} studied the observed VLA radio fluxes of Cyg~OB2 No.~9 at 3.6, 6, and 20 cm. They found a 2.35 yr period from the data. A preliminary 6 cm light curve from our modelling shows variability in the radio flux linked to the orbital period that is the fingerprint of non-thermal radiation. Compared to the observations, the theoretical fluxes are much too high and the maximum occurs too early.

\section{Future work}
To improve the current results for Cyg~OB2 No.~8A and 9, we need to include the orbital motion and solve the hydrodynamics equations. A more detailed study is also necessary to better determine the star and wind parameters of the binary components and to investigate the porosity/clumping problem.


\begin{thebibliography}{}

\bibitem[Adam 1990]{Adam90}
{Adam, J.} 1990,
\textit{A \& A}, 240, 541 

\bibitem[Antokhin et al. 2004]{Antokhin_etal04}
{Antokhin, I.I., Owocki, S.P., \& Brown, J.C.} 2004, 
\textit{ApJ}, 611, 434

\bibitem[De Becker et al. 2006]{DeBecker_etal06}
{De Becker, M., Rauw, G., Sana, H., Pollock A.M.T., Pittard, J.M., Blomme, R., Stevens, I.R., \& van Loo, S.} 2006, 
\textit{MNRAS}, 371, 1280

\bibitem[Blomme et al. 2010]{Blomme_etal10}
{Blomme, R., De Becker, M., Volpi, D. \& Rauw, G.} 2010,
\textit{arXiv 1006.3540}, accepted by A \& A

\bibitem[Eichler \& Usov (1993)]{EichlerUsov93}
{Eichler, D. \& Usov, V.} 1993, 
\textit{ApJ}, 402, 271

\bibitem[Martins et al. 2005]{Martins_etal05}
{Martins, F., Schaerer, D., \& Hillier, D.~J.} 2005,
\textit{A \& A}, 436, 1049

\bibitem[Naz\'e et al. 2010]{Naze_etal10}
{Naz\'e, Y., Damerdji, Y., Rauw, G., Kiminki, D.~C., Mahy, L., Kobulnicky, H.~A., Morel, T., De Becker, M., Eenens, P. \& Barbieri, C.} 2010,
\textit{ApJ}, 719, 634

\bibitem[Van Loo et al. 2008]{vanLoo_etal08}
{van Loo, S., Blomme, R., Dougherty, S.~M., \& Runacres, M.~C.} 2008,
\textit{A \& A}, 483, 585

\bibitem[Vink et al. 2001]{Vink_etal01}
{Vink, J.~S., de Koter, A. \& Lamers, H.~J.~G.~L.~M.} 2001,
\textit{A \& A}, 369, 574


\end{thebibliography}
\end{document}